\documentclass[onecolumn,trackchanges]{aastex631}

\usepackage{amsmath}
\usepackage{xcolor}
\usepackage{graphicx}
\usepackage{float}
\usepackage{subfigure}
\usepackage{hyperref}
\usepackage{comment}
\usepackage{url}
\begin{document}

\title{Shear Particle Acceleration in Structured Gamma-Ray Burst Jets: III. The Radiation Physics of Bright Prompt Optical Flash}\

\correspondingauthor{Xiao-Li Huang, En-Wei Liang}
\email{xiaoli.huang@gznu.edu.cn, lew@gxu.edu.cn}

\author[0009-0001-8025-3205]{Zi-Qi Wang}
\affiliation{Guangxi Key Laboratory for Relativistic Astrophysics, School of Physical Science and Technology, Guangxi University, Nanning 530004, People’s Republic of China}

\author[0000-0002-9725-7114]{Xiao-Li Huang}
\affiliation{School of Physics and Electronic Science, Guizhou Normal University, Guiyang 550025, People’s Republic of China}

\author[0000-0002-7044-733X]{En-Wei Liang}
\affiliation{Guangxi Key Laboratory for Relativistic Astrophysics, School of Physical Science and Technology, Guangxi University, Nanning 530004, People’s Republic of China}

\begin{abstract}
The radiation physics of bright prompt optical emission of gamma-ray bursts (GRBs) remains a puzzle. Assuming that the GRB ejecta is structured, we investigated this issue by characterizing the ejecta as an ultra-relativistic uniform jet core surrounded by a mild-relativistic cocoon. The mixed jet-cocoon (MJC) region can accelerate particles through the shear acceleration mechanism. Parameterizing the radial velocity profile of the MJC region with an exponential function and assuming a uniform magnetic field configuration, we show that the synchrotron radiation of the shear-accelerated electrons can produce a bright optical flash. Emission of the self-synchrotron Compton (SSC) process of the electron population can result in an X-ray excess and an extra MeV-GeV gamma-ray flash relative to the Band function component in the keV-MeV band, which is attributed to the synchrotron radiation of the shock-accelerated electrons in the jet core. Our model reasonably represents the extremely bright optical flash and spectral characteristics of GRBs 990123, 080319B, and 130427A. The inferred magnetic field strength of the MJC region is above $10^{5}$ G, potentially suggesting that the jets of these GRBs are highly magnetized.

\end{abstract}


\section{Introduction} \label{sec:intro}
Gamma-ray bursts (GRBs) are an extreme astrophysical phenomenon resulting from the collapse of massive stars or the merger of compact binary objects \citep{1986ApJ...308L..43P,1992ApJ...395L..83N,1993ApJ...405..273W,1998Natur.395..670G,2005Natur.437..851G}. 
The prompt gamma-ray emission spectra in the keV-MeV band of most GRBs observed with the Burst And Transient Source Experiment (BATSE) on board the Compton Gamma Ray Observatory (CGRO) and the Gamma-Ray Burst Monitor (GBM) on board the {\em Fermi} mission are typically non-thermal and can be empirically described by the Band function, which is a smoothly broken power-law function depicted as $N\left( E \right) \propto E^{- \alpha}$ at low energy band and $N\left( E \right) \propto E^{- \beta}$ at high energy band \citep{1993ApJ...413..281B,2006ApJS..166..298K}.  
Interestingly, broadband observations reveal distinct components over the Band function in some GRBs. For example, the {\em Fermi} mission discovers a bimodal feature in the keV-MeV-GeV band in several GRBs, such as GRBs 090926A, 110731A, and 240825A \citep{2011ApJ...729..114A,2013ApJ...763...71A,2025ApJ...984L..45Z}. The spectrum of GRB 090902B is characterized by a quasi-thermal emission spectrum plus a single power-law component with spectral coverage from the X-ray to the GeV band \citep{2009ApJ...706L.138A,2010ApJ...709L.172R}. At the low energy end, a soft X-ray excess (approximately from a few keV to tens of keV) is seen in a small fraction ($\sim 15\%$) of GRBs observed with {\em CGRO}/BATSE \citep{1996ApJ...473..310P,2000ApJS..126...19P,2006A&A...447..499V} and with the {\em Swift}/Burst Alert Telescope (BAT; \citealp{2014ApJ...795..155P}). In addition, bright prompt optical flashes were observed in GRB 990123 \citep{1999Natur.398..400A}, GRB 080319B \citep{2008Natur.455..183R}, and GRB 130427A \citep{2014Sci...343...38V}. The optical flux is not a simple power-law extension of the soft X-ray to the optical band. The relation between prompt optical flashes and gamma-ray emission has been explored to assess whether they are physically related \citep{2005Natur.435..178V,2006Natur.442..172V,2008Natur.455..183R}. Statistical analyses with a limited sample suggest that they are likely distinct components \citep{2005Natur.435..178V,2012ApJ...753L..31A,2016ApJ...818..190F,2018ApJ...859...70F}.
These broadband observations challenge the conventional internal shock model of GRBs, in which the prompt emission is attributed to the synchrotron radiation and/or the synchrotron self-Compton (SSC) process of relativistic electrons accelerated in the jet by internal shocks \citep{1994ApJ...430L..93R,1996ApJ...466..768T,1999ApJ...512..699C,2001A&A...372.1071D,2011A&A...526A.110D} or by magnetic reconnection 
\citep{1998MNRAS.296..275D,2003astro.ph.12347L,2009A&A...498..677B,2011ApJ...726...90Z} \footnote{Alternatively, in the matter-dominated fireball scenario, photospheric thermal emission can also make a contribution in the keV-MeV band \citep{1986ApJ...308L..43P,1986ApJ...308L..47G,1994MNRAS.270..480T,2005ApJ...628..847R,2006ApJ...642..995P,2013MNRAS.428.2430L}.}. 
\cite{2009ApJ...692.1662Y} proposed that a model of two-component synchrotron emission from internal shocks can explain prompt optical and gamma-ray emission of GRB 080319B. Through Monte Carlo simulations, \cite{2010ApJ...725L.121A}  modeled the low-energy spectral excess components as hadronic cascade emission initiated by photomeson interactions of ultra-high-energy protons accelerated within GRB outflows. These models involve an extra distinct emitting electron or proton population.      

Relativistic numerical simulations reveal the presence of a jet-cocoon structure of the GRB ejecta when it propagates through a dense ambient medium  \citep{2000ApJ...531L.119A,2004ApJ...608..365Z,2013ApJ...767...19L,2022ApJ...933L...9G,2022ApJ...933L...2G,2023MNRAS.520.1111H}. The structured GRB jet model has been studied \citep{2003ApJ...595L..33S,2005ApJ...626..966P,2017ApJ...848L...6L,2018MNRAS.478.4553D,2018MNRAS.479..588G,2023MNRAS.522L..56S}. Most importantly, the observations of GRB 170817A present convincing evidence for a structured jet of this particular GRB \citep{2018Natur.554..207M,2018ApJ...863L..18A,2019Sci...363..968G}. The material mixing transition region between the jet and the cocoon exhibits a continuous velocity profile. Such a mixed jet-cocoon (MJC) region serves as a promising site for particle acceleration via shear flows
\citep{1981SvAL....7..352B,1990ICRC....4..126W,1998A&A...335..134O,2004ApJ...617..155R,2005ApJ...632L..21R,2018ApJ...855...31W,2021ApJ...907L..44S,2023ApJ...958..169W}.         
We showed that, through synchrotron radiation and/or SSC process, the emission of the shear-accelerated electrons, together with the electron population accelerated by internal shocks or magnetic reconnection in the jet core can account for the bimodal feature of the prompt gamma-ray spectrum and the observations of GRB 170817A  (\citealp{2024ApJ...977..182W,2025ApJ...981..196W}; Paper I and Paper II of this series). As the third paper of this series, we investigate whether the broadband prompt emission spectrum in the optical, X-ray, and gamma-ray bands can be explained by the proposed model.

We select three GRBs that have bright prompt optical flares, i.e., GRBs 990123 \citep{1999Natur.398..400A}, 080319B \citep{2008Natur.455..183R}, and 130427A \citep{2014Sci...343...38V}, as examples for our analysis. 
The structure of this article is outlined as follows. Sec.~\ref{sec:Shear acceleration} presents our model and demonstrates the predicted broadband spectra. Sec.~\ref{sec:Case Study} presents the application of our model to interpret the broadband prompt emission spectra of the three GRBs. The summary and discussions are presented in Sec.~\ref{sec:conclusionanddiscussion}. In this paper, we employ a Hubble constant of $H_0=71\ \mathrm{km} \mathrm{s}^{-1}\,\,\mathrm{Mpc}^{-1}$, and the cosmological parameters of $\varOmega _M=0.27$ and $\varOmega _{\Lambda}=0.73$.

\section{Model} \label{sec:Shear acceleration}
As illustrated in Paper I, we assume that the GRB ejecta consists of a uniform jet core (with an opening angle $\theta_{\rm jet}$) surrounded by a mild-relativistic cocoon (with an opening angle $\theta_{\mathrm{cn}}$). In conventional internal shock GRB models, the dissipation radius of prompt emission is commonly inferred around $10^{13}\sim 10^{15}$ cm \citep{2004RvMP...76.1143P,2010MNRAS.407.1033B}. In the present calculations, the radius of the jet core is $R=10^{15}$ cm. The boundaries of the MJC region are defined as $ r_0 <r <r_2$, where $r_0=R\theta_{\rm jet} / 2$ and $r_2=R\theta_{\rm cn} / 2$. We set $\theta_{\mathrm{jet}} = 0.07$ rad and $\theta_{\rm cn}=0.7$ rad.

Electrons in the jet core are accelerated via internal shocks through the first-order Fermi acceleration mechanism. We assume an isotropic pitch-angle distribution of the electrons. The distribution of the accelerated electrons is empirically represented by a broken power-law as \citep{1996ApJ...473..204S,1998ApJ...497L..17S,2018pgrb.book.....Z}
\begin{equation}
\frac{dN_{e,\mathrm{jet}}}{d\gamma _{e,\mathrm{jet}}}\propto \begin{cases}
	\gamma _{e,\mathrm{jet}}^{-2}&		\gamma _{\mathrm{m},\mathrm{jet}}\leqslant \gamma _{e,\mathrm{jet}}\leqslant \gamma _{\rm b,jet}\\
	\gamma _{e,\mathrm{jet}}^{-p_{\mathrm{jet}}-1}&		\gamma _{\rm b,jet}<\gamma _{e,\mathrm{jet}}\leqslant \gamma _{\rm M,jet}\\
\end{cases},  
\end{equation}
where $\gamma_e$ is the electron Lorentz factor, $p_{\rm jet}$ is the spectral index and $\gamma_{\rm m,jet}$, $\gamma_{\rm b,jet}$, and $\gamma_{\rm M,jet}$ are characteristic Lorentz factors corresponding to the minimum, break, and maximum energies of the electrons, respectively. This simplification is motivated by two considerations: (1) Prompt emission is typically assumed to occur in the fast-cooling regime of synchrotron emission, where the Fokker-Planck equation approximately leads to a broken power-law distribution \citep{2018pgrb.book.....Z}; (2) A simple power or broken power-law can qualitatively represent electron distributions inferred from observational fitting, reasonably consistent with theoretical expectations from shock acceleration \citep{2004ApJ...613..460B,2014ApJ...784...17B}.

\begin{figure}[htbp!]
    \centering
    \includegraphics[width=0.35\textwidth]{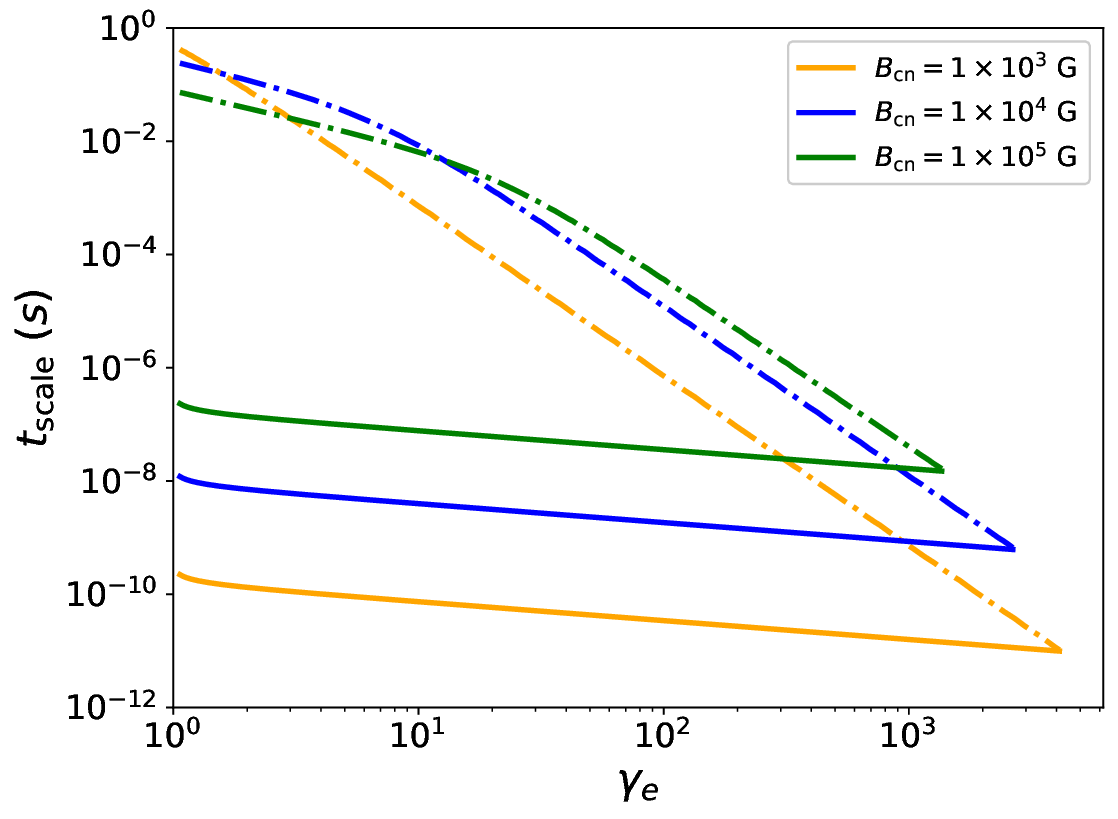} 
    \includegraphics[width=0.35\textwidth]{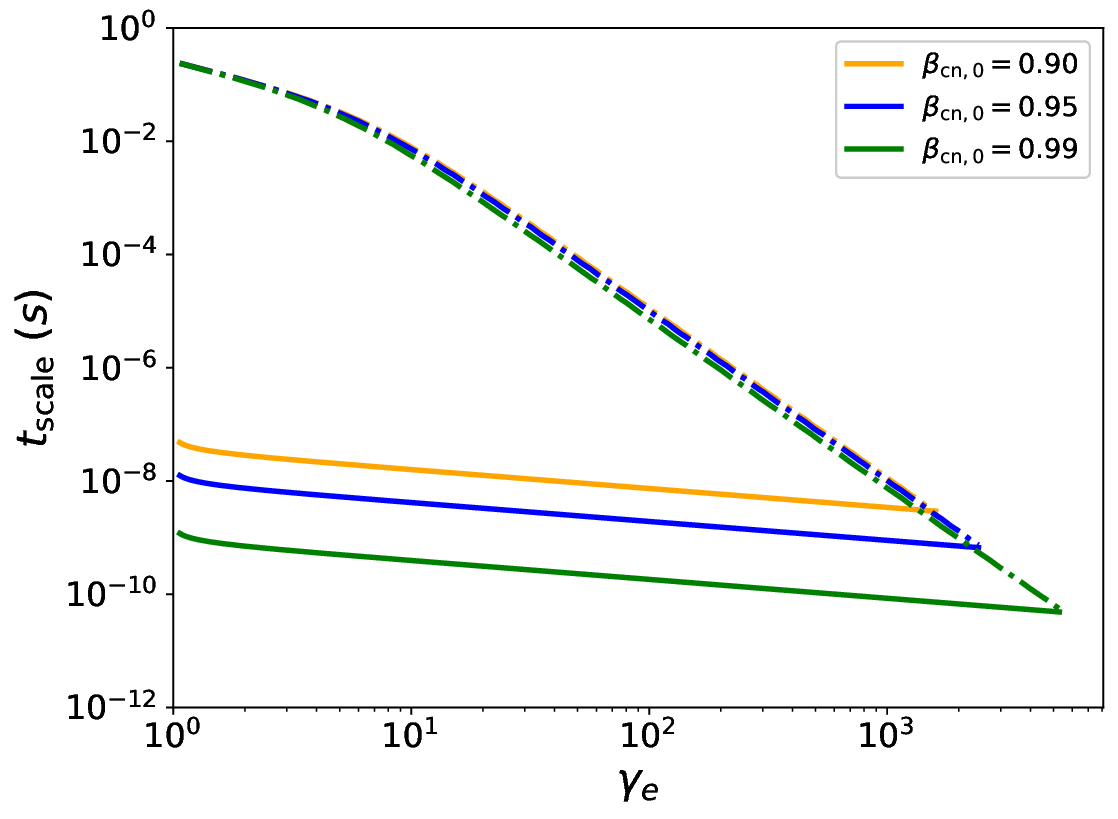}
    \includegraphics[width=0.35\textwidth]{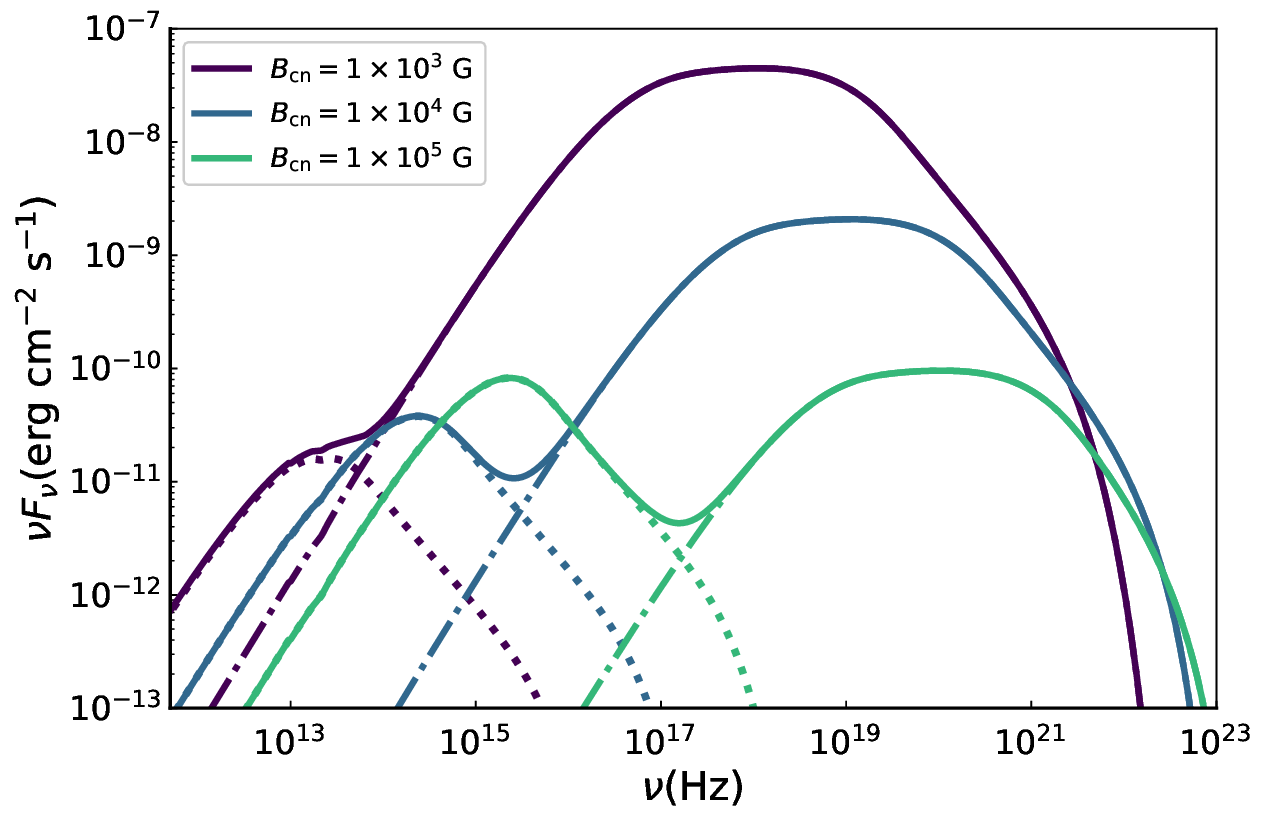}
    \includegraphics[width=0.35\textwidth]{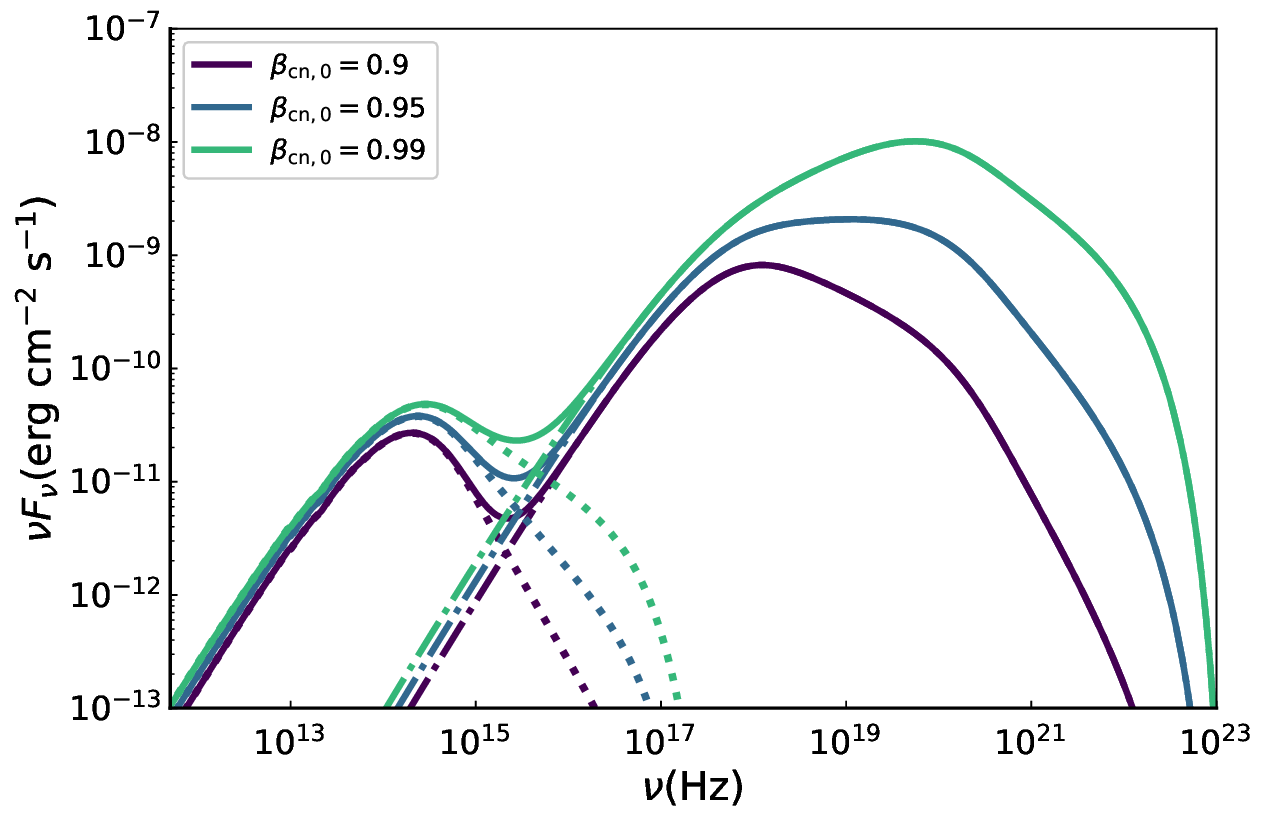}
  
    \caption{{\em Upper panels---}Acceleration and radiative cooling timescales of shear-accelerated electrons as a function of $\gamma_e$ computed for various combinations of $B_{\rm cn}$ and $\beta_{\rm cn,0}$ as indicated. {\em Lower panels---}Spectral energy distributions of synchrotron radiation and SSC process emission of the shear-accelerated electrons.}
    \label{fig:time_SED}
\end{figure}

Electrons with initial momentum $p_0 = \gamma_{e, \mathrm{inject}} m_e c$ are injected into the MJC region at $r = r_1\ (r_1 \gtrsim r_0)$, and experience further acceleration through the shear acceleration mechanism. 
The fluid velocities at the inner and outer boundaries in units of the speed of light are $\beta_{\rm cn,0}$ and $\beta_{\rm cn,2}$, respectively. The radial velocity profile is described by an exponential function $u_{\rm cn}(r)=\beta _{\rm cn,0}e^{-k}$, where $k = r \ln (\beta_{\rm cn,0}/\beta_{\rm cn,2})/{r_{2}}$ \citep{2024ApJ...977..182W,2025ApJ...981..196W}. 
The particle distribution ($f_{0}$) for shear acceleration in relativistic shear flows follows the transport equation. In the strong scattering regime and assuming that the scattering wave frame is aligned with the comoving fluid frame, the analytical electron distribution function $f_0$ is formulated as \citep{2018ApJ...855...31W}
\begin{equation}
\begin{split}
    f_0=&\frac{15}{8 \pi^{2} \left( \xi _0-\xi _2 \right)\left| \frac{d\xi}{dr}|_{r_1} \right|r_1}\left( \frac{N_0}{{p_0}^{3}c^2\tau _0} \right) \exp \left[ -\frac{\left( 3+\alpha \right) T}{2} \right] \\
    &\times \sum_{n=0}^{\infty}{\frac{1}{y_n}\sin \left[ \left( n+\frac{1}{2} \right) \pi w_1 \right]}\sin \left[ \left( n+\frac{1}{2} \right) \pi w \right] \exp \left( -y_n\left| T \right| \right) ,
\label{eq:f0}
\end{split}
\end{equation}
 
\begin{equation}
\xi \left( r \right) =\frac{1}{2}\ln \left( \frac{1+u_{\mathrm{cn}}}{1-u_{\mathrm{cn}}} \right) , \ \  w \equiv \frac{\xi -\xi _2}{\xi _0-\xi _2}, \ \ T=\ln \left( \frac{p}{p_0} \right), 
\label{eq:6}
\end{equation}
and
\begin{equation}
y_n=\left[ \frac{5\pi ^2\left( 2n+1 \right) ^2}{4\left( \xi _0-\xi _2 \right) ^2}+\frac{\left( 3+\alpha \right) ^2}{4} \right] ^{{{1}/{2}}}, \ \  n=0,1,2,\dots ,
\label{eq:7}
\end{equation}
where $\tau_0$ is the initial scattering timescale and $\alpha = 2 - q$ characterizes the momentum dependence of the mean scattering time. We adopt $\alpha=1/3$, corresponding to $q=5/3$, which is consistent with the Kolmogorov turbulence model \citep{1941DoSSR..30..301K,2002ApJ...564..291C}. The subscripts 0, 1, and 2 correspond to the quantities in $r=r_0$, $r=r_1$, and $r=r_2$, respectively. Acceleration timescale is estimated by \citep{2017ApJ...842...39L,2018ApJ...855...31W}
\begin{equation}
t_{\mathrm{acc},\mathrm{cn}}=\frac{15}{\left(4+\alpha \right) {\Gamma^4_{\mathrm{cn}}}\left( du_{\rm cn}(r)/dr \right) ^2\tau} ,
\end{equation}
where $\Gamma_{\rm cn}$ is the Lorentz factor of the MJC region and $\tau$ is the mean scattering time. 
Accelerated electrons in the MJC region undergo energy losses through $\rm Syn$ and $\rm SSC$ processes. Our calculations assume a fully random magnetic field configuration and the electron emission is isotropic in the comoving frame \citep{1986A&A...164L..16C}. The maximum electron Lorentz factor $\gamma_e$ is determined by the condition $t_{\rm acc,cn}=t_{\rm cool}$, where $t_{\rm cool}$ is the radiative cooling timescale arising from both $\rm Syn$ and $\rm SSC$ processes, and expressed as \citep{1996ApJ...473..204S,2009ApJ...703..675N}
\begin{equation}
t_{\rm cool}=t_{\rm Syn}+t_{\rm SSC}=\frac{6\pi m_ec}{\gamma _{e}\sigma_T B^2\left( 1+ Y \right)} ,
\end{equation}
where $\sigma_T$ is the Thomson cross section and $Y$ is the Compton parameter defined as the ratio of the SSC power to the Syn power. 
In the upper panels of Figure~\ref{fig:time_SED}, we present the shear acceleration and cooling timescales as a function of the electron Lorentz factor $\gamma_e$, with $\gamma_{e,\mathrm{inject}} = 100$. 
The upper-left panel illustrates the case in which $B_{\rm cn}$ is varied ($10^3$, $10^4$, and $10^5$ G) by fixing $\beta_{\rm cn,0}=0.90$. The upper-right panel presents results for varying $\beta_{\rm cn,0}$ (0.90, 0.95, and 0.99) by fixing $B_{\rm cn} = 10^4$ G. In both cases, the maximum Lorentz factor of shear-accelerated electrons is $\gamma_{\rm M, cn} \sim 10^3$.
The lower panels of Figure~\ref{fig:time_SED} show the spectral energy distributions (SEDs) emitted by the shear-accelerated electron population for different sets of $\{B_{\rm cn}, \beta_{\rm cn,0}\}$. The synchrotron emission characteristically peaks in the infrared to optical band. The peak frequency ($\nu_{p, \rm syn}$) of synchrotron radiations is not sensitive to $\beta_{\rm cn,0}$ at a given $B_{\rm cn}$, but shifts toward higher energies with a corresponding decrease in peak flux as $B_{\rm cn}$ increases. The peak flux level of the $\rm SSC_{cn,0}$ component strongly depends on $\beta_{\rm cn}$. In the scenario of $\beta_{\rm cn,0}=0.99$, a bright GeV flash accompanied by the optical flash is expected to be detectable.

\section{Case study}\label{sec:Case Study}
\begin{figure}[htbp!]
    \centering
    \includegraphics[width=0.32\textwidth]{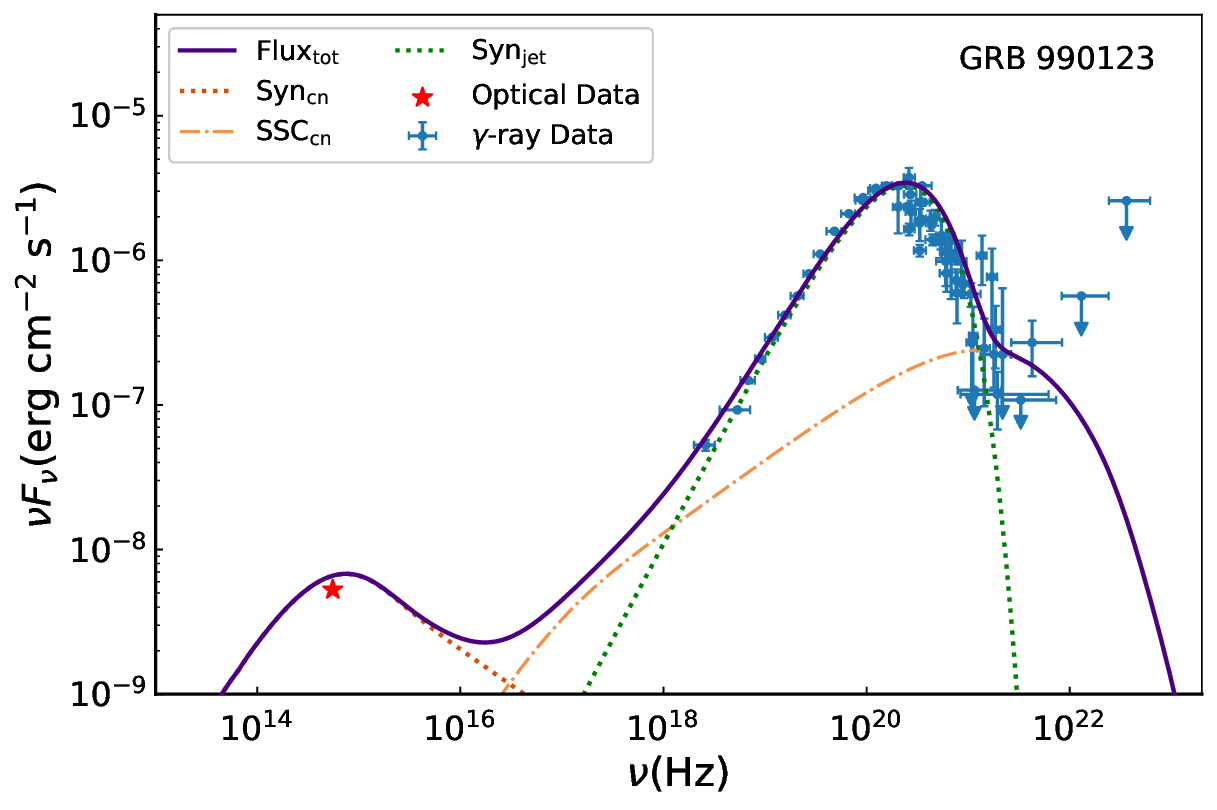}
    \includegraphics[width=0.32\textwidth]{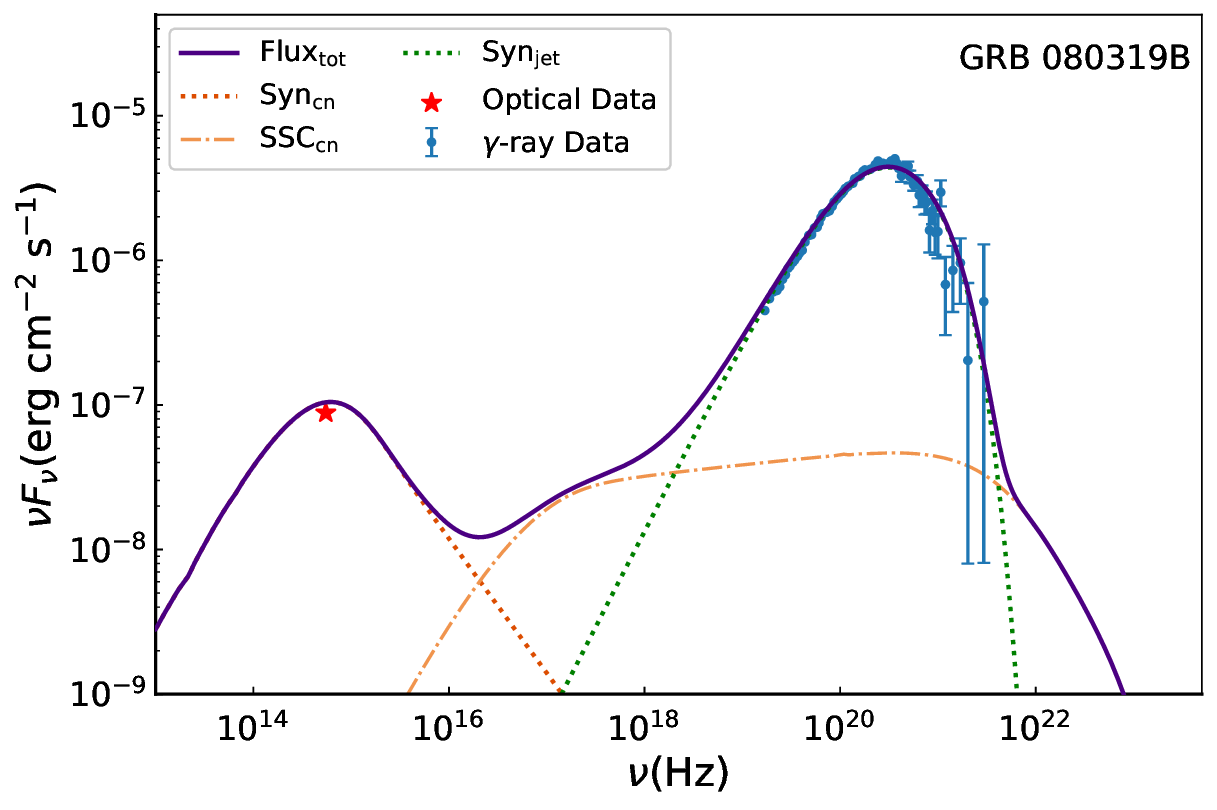}
    \includegraphics[width=0.32\textwidth]{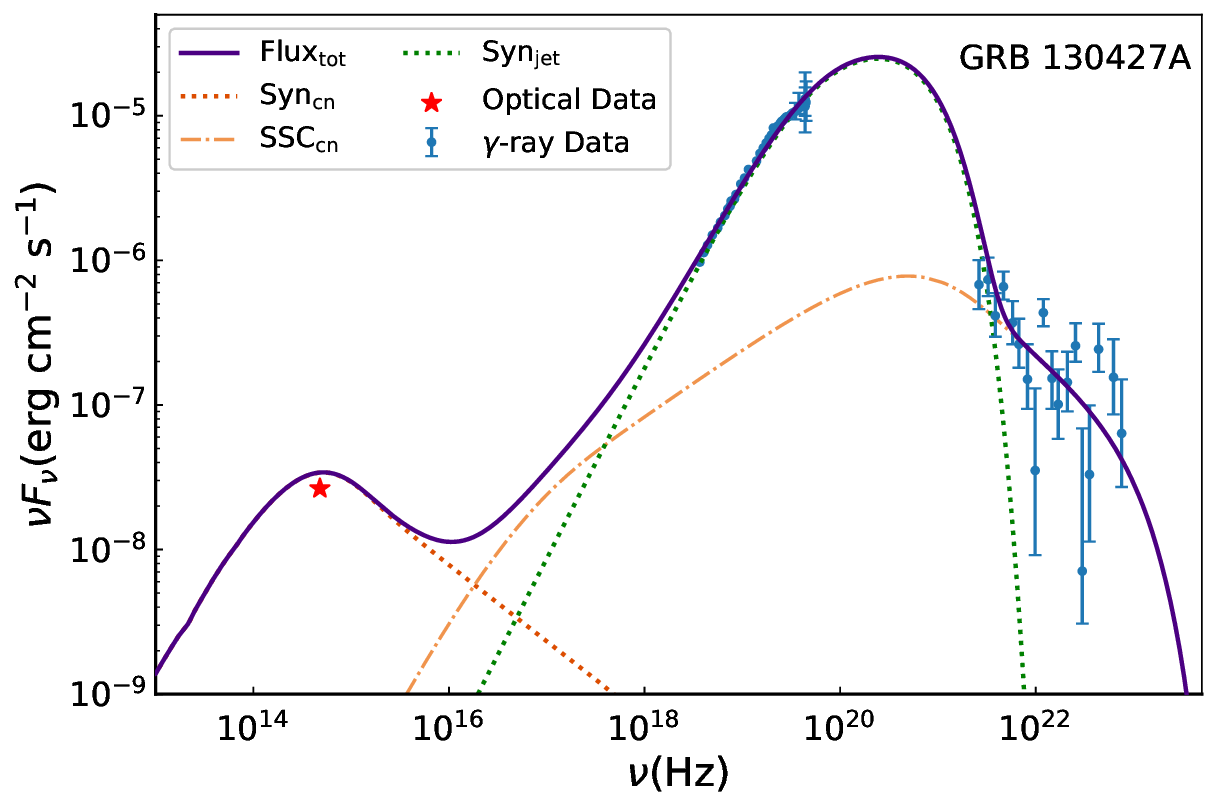}
    \caption{Prompt emission spectra (data points) of GRBs 990123, 080319B, 130427A in the optical, X-ray, and gamma-ray bands. Solid lines represent the theoretical model curves, and dashed lines indicate different emission components from the MJC region and jet core.}
    \label{fig:case}
\end{figure}

We apply our model to explain the broadband prompt emission spectra of GRBs 990123, 080319B, and 130427A. The data together with the fits with our model are presented in Figure~\ref{fig:case}. The model-predicted optical flux has been corrected for Galactic dust extinction \citep{2003ARA&A..41..241D}. The model parameters are summarized in Table~\ref{tab:parameters}. 

\begin{itemize}
    \item GRB 990123. It is an extremely bright GRB at $z = 1.60$ \citep{1999Sci...283.2075A}. Its prompt emission was simultaneously detected in the optical, X-ray, and gamma-ray bands \citep{1999Natur.398..400A,1999IAUC.7100....1A,1999ApJ...524...82B}. Although the time-integrated prompt gamma-ray spectrum observed with CGRO/BATSE in the keV-MeV band can be fitted with the Band function, an X-ray excess below $\sim 15$ keV over the Band function component is found in some slices of its time-resolved spectra \citep{1999ApJ...524...82B}. Additionally, a MeV-GeV excess over that Band function is also observed with a low significance level. The fluxes of the gamma-ray and optical emission are not correlated, and the optical flux cannot be simply extrapolated from the gamma-ray spectrum to the optical band \citep{1999ApJ...524...82B}. These spectral features are likely indicators of another distinct emission component \citep{1999ApJ...524...82B,2005A&A...438..829C}. 
    Although the optical data and sub-MeV gamma-ray emission can be explained as synchrotron and SSC emission with the standard internal shock model, the predicted MeV-GeV flux by the second-order SSC process of this scenario cannot explain the flux limits at high energy range and the tentative MeV excess \citep{1999ApJ...518L..73F,2007MNRAS.376.1065P}. 
    We apply our model to the broadband spectrum of GRB 990123. As shown in Figure~\ref{fig:case}, the model effectively reproduces the observed spectrum by attributing the prompt optical and high-energy gamma-ray components to synchrotron and SSC radiation of the shear-accelerated electrons, respectively. The weak soft X-ray excess is also contributed by the $\rm SSC_{\rm cn}$ component. The prompt gamma-ray peak at several hundreds of keV is overwhelmingly dominated by the synchrotron radiation of the shock-accelerated electrons in the jet core.

    \item GRB 080319B. This is the well-known ``naked-eye" burst, with a redshift of $z=0.937$ \citep{2008GCN..7444....1V,2008Natur.455..183R}. 
    The burst was triggered by the {\em Swift}/BAT (15-150 keV) and the Konus/Wind (18 keV-12 MeV) \citep{2008GCN..7427....1R,2008GCN..7482....1G}. 
    The wide-field robotic instrument Telescopio Ottimizzato per la Ricerca dei Transienti Ottici RApidi (TORTORA) observed the brightest segment of the optical flash with high temporal resolution \citep{2009ApJ...691..723B}. We use the optical and gamma-ray data compiled by \cite{2012ApJ...753L..31A} from the Konus/Wind and TORTORA observations. They are illustrated in Figure~\ref{fig:case}.       
    As proposed by \cite{2008Natur.455..183R} and \cite{2008MNRAS.391L..19K}, although the optical and keV-MeV gamma-ray emission can be explained by the synchrotron radiation plus SSC emission, the predicted GeV gamma-ray flux from the second-order inverse Compton (IC) scattering in this scenario exceeds the observational limits in the MeV band by approximately a factor of 10. 
    \cite{2009ApJ...692.1662Y} interpreted the prompt optical and keV-MeV gamma-ray emission as synchrotron radiation of two distinct electron populations. \cite{2010ApJ...725L.121A} modeled the prompt optical and X-ray excess components as hadronic cascade emission of ultra-high-energy protons accelerated within the GRB outflow, but the search for high-energy muon neutrinos from GRB 080319B with the IceCube Neutrino Telescope does not support this scenario \citep{2009ApJ...701.1721A}. The prompt emission spectrum of GRB 080319B can be well represented with our model, as shown in  Figure~\ref{fig:case}. Similar to GRB 990123, the bright prompt optical emission is attributed to the synchrotron radiation of the shear-accelerated electrons. 
    The model further predicts an X-ray excess below 10 keV arising from the $\rm SSC_{cn}$ component of the shear-accelerated electrons. Compared with GRB 990123A, the high-energy gamma-ray flux from the $\rm SSC_{\rm cn}$ component is obviously suppressed. This agrees with the observed high-energy gamma-ray flux above $10$ MeV.

    \item GRB 130427A. This is an extremely powerful GRB located at $z=0.34$ \citep{2013GCN.14455....1L,2014Sci...343...42A}.     
    A prompt GeV flash with photon energy reaching up to 73 GeV was detected with {\em Fermi}/LAT \citep{2014Sci...343...42A,2018ApJ...859...70F}. Meanwhile, a bright optical flash with a peak brightness of $R=7.03\pm0.03$, temporally correlated with the GeV flash, was also detected \citep{2014Sci...343...38V}. We take its broadband data from \cite{2014Sci...343...42A} and \cite{2018ApJ...859...70F}. The SED is characterized by a Band function component in the keV–MeV range with a pronounced excess in the GeV band, similar to GRB 990123.
    \cite{2014ApJ...788...36B} proposed that the optical flash arises from synchrotron emission of a thermal plasma behind the forward shock, whereas the GeV flash originates from the IC scattering process of the electrons in the plasma. Based on TeV gamma-ray flux upper limits reported by Milagro and HAWC observations, \cite{2018ApJ...859...70F} suggested that the optical and GeV gamma-ray flashes are produced by a distinct electron population. By attributing the optical and GeV flashes to the synchrotron radiation and the SSC process emission of shear-accelerated electrons, our model adequately reproduces the SED data as illustrated in Figure~\ref{fig:case}. The predicted high-energy gamma-ray flux is consistent with the observed prompt gamma-ray photons with energy extending up to 73 GeV.
\end{itemize}
As summarized in Table~\ref{tab:parameters}, the model parameters derived from our fits are comparable among the three GRBs, i.e. $\beta_{\rm cn,0} =0.963\sim 0.994$, $ B_{\rm cn}=(1.10\sim 3.10) \times 10^{5}$ G, $\gamma_{e,\rm{inject}}=9.6\sim 11$, and $\gamma_{\rm M, cn}=(1.48\sim 1.94)\times 10^{3}$ for the MJC region and $\Gamma_{\rm jet}=450\sim 500$, $B_{\rm jet}=(4\sim 5)\times 10^{5}$ G, $p_{\rm jet}=2.15\sim 2.35$, $\gamma_{\rm m, jet}=(4.1\sim 8.76)\times 10^{2}$, and $\gamma_{\rm b, jet}=(1.17\sim 1.97)\times 10^3$, and  $\gamma_{\rm M, jet}=(1.31\sim 2.02)\times 10^{3}$ for the jet core. These results suggest that the MJC region is moderately relativistic, with a strong magnetic field strength comparable to that of the jet core.

\begin{table}[htbp!]
    \centering
    \caption{{Model parameters derived from joint spectral analyses of optical and gamma-ray observations.}}
    \begin{tabular}{ccccc|cccccc}
        \hline\hline
        Name & $\beta _{\mathrm{cn},0}$ & $B_{\mathrm{cn}}$ & $\gamma _{e, \mathrm{inject}}$ & $\gamma_{\rm M,cn}$ &$\Gamma_{\rm jet}$ & $B_{\rm jet}$ & $p_{\rm jet}$ & $\gamma_{m, \rm jet}$ & $\gamma_{\rm b,jet}$ & $\gamma _{\rm M, jet}$ \\
        \hline
        GRB 990123  & $0.994$ & $2.94 \times 10^{5} $ & $11.0$ & $1.48 \times 10^{3}$ & 450 & $4 \times 10^{5}$ & $2.35$ & $8.76 \times 10^{2}$ & $1.17 \times 10^{3}$ & $1.31 \times 10^{3}$ \\
        \hline
        GRB 080319B & $0.963$ & $3.10 \times 10^{5} $ & $9.6$  & $1.74 \times 10^{3}$ &500 & $5 \times 10^{5}$ & $2.15$ & $7.70 \times 10^{2}$ & $1.97 \times 10^{3}$ & $2.02 \times 10^{3}$ \\
        \hline
        GRB 130427A & $0.990$ & $1.10 \times 10^{5} $ & $10.3$ & $1.94 \times 10^{3}$ &455 & $5 \times 10^{5}$ & $2.20$ & $4.10 \times 10^{2}$ & $1.65 \times 10^{3}$ & $1.73 \times 10^{3}$\\
        \hline
    \end{tabular}
    \label{tab:parameters}
\end{table}  

\section{Conclusions and Discussion} \label{sec:conclusionanddiscussion}
Characterizing the structured ejecta of GRBs as an ultra-relativistic uniform jet core surrounded by a mild-relativistic cocoon, we have shown that the prompt optical, X-ray, and gamma-ray emission can be represented by the synchrotron radiation and the SSC process of both the shear-accelerated electrons in the MJC region and the shock-accelerated electrons in the jet core. Describing the radial velocity profile of the MJC region with an exponential function and assuming a uniform magnetic field configuration, the injected electron with $\gamma_e\sim 10$ can be accelerated up to $\gamma_{e,\max}\sim 10^3$ through shear acceleration if $\beta_{\rm cn,0}=0.9\sim 0.99$ and $B_{\rm cn}\sim 10^{5}$ G. The synchrotron and SSC emissions of these electrons can produce a bright optical flash, accompanied by a MeV-GeV gamma-ray flash, as well as an X-ray excess relative to the conventional Band function component, which is attributed to the synchrotron radiation of the shock-accelerated electrons in the jet core. Our model reasonably represents the broadband prompt emission SEDs of GRBs 990123, 080319B, and 130427A, the three optically brightest GRBs observed to date. 

The model parameters derived from our spectral fits are comparable among the three GRBs, i.e. $\beta_{\rm cn,0} >0.9$, $ B_{\rm cn}\sim 10^{5}$ G for the MJC region and $\Gamma_{\rm jet}\sim 500$, $B_{\rm jet}\sim 5\times 10^{5}$ G for the jet core. The magnetic field strength of the MJC regions is comparable to the jet core. These bursts are associated with the collapse of massive stars. The strong magnetic field of the MJC region could result from extremely powerful explosions through a dense stellar wind medium. This involves complex mechanisms of magnetic field generation and amplification, including the Weibel instability at the shock front, the Kelvin-Helmholtz instability at shear interfaces, the Rayleigh-Taylor instability in mixing layers, the two-stream instability, and other related mechanisms \citep{1961hhs..book.....C,1995ApJ...453..332J,2015NatPh..11..173H,2021PhRvL.126u5101P}.
The injected electrons inferred from the model have a factor of $\gamma_{\rm e, inject} \sim 10$. They may be jet-head electrons heated by shock waves as the jet propagates through the dense envelope and injected to the MJC region during radial diffusion, or background electrons energized by local turbulence and/or magnetic reconnection within the MJC region \citep{2011ApJ...731...80N,2013ApJ...766L..19L,2020PhR...866....1L}.

\begin{figure}[htbp!]
    \centering
    \includegraphics[width=0.35\textwidth]{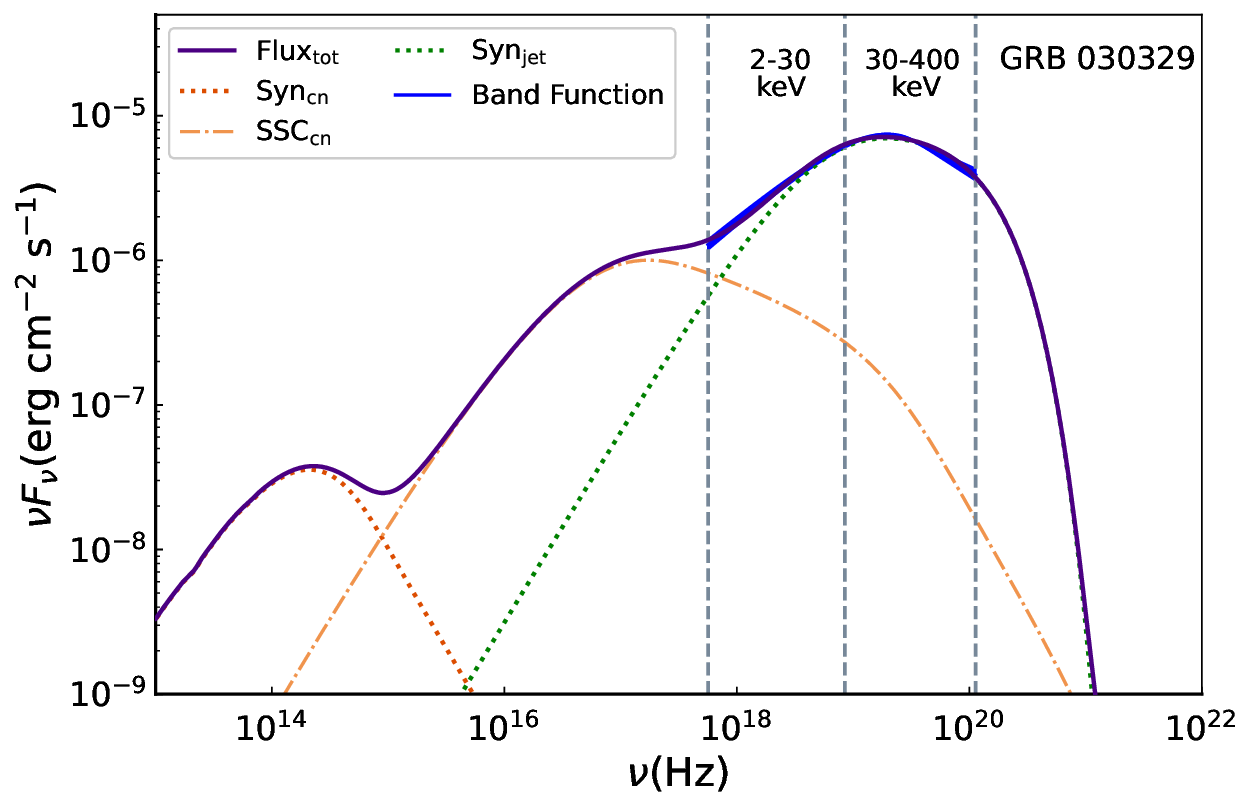}
    \caption{The best-fit of the prompt gamma-ray data of GRB 030329 with the Band function (the blue shadow, taken from\citealp{2004ApJ...617.1251V}) together with the fit curve with our model (solid line). The dashed and dotted lines mark different emission components in the model.}
    \label{fig:030329}
\end{figure}

The SSC process emission of the shear-accelerated electrons in the MJC region may shape an excess in the soft X-ray band (1-10 keV) over the synchrotron radiation of the shock-accelerated electrons in the jet core, as shown in Figure~\ref{fig:case}. 
Notably, an additional X-ray component was tentatively identified in the X-ray-rich burst GRB 030329. Its X-ray lightcurve (2-30 keV) does not trace the temporal behavior of the gamma-ray lightcurve (30-400 keV) \citep{2004ApJ...617.1251V}. 
Although its prompt gamma-ray spectrum can be described with the Band function, the low-energy spectral index of time-resolved spectra evolves from $-0.89$ to $-1.90$, likely indicating the presence of a distinct X-ray emission component \citep{2004ApJ...617.1251V}. We fit the prompt gamma-ray spectrum with our model. The derived parameters are $\beta _{\mathrm{cn},0}=0.918$, $B_{\mathrm{cn}}=9.0 \times 10^4$ G, $\gamma _{e, \mathrm{inject}}=12.7$, $\Gamma_{\rm jet}=300$, $B_{\rm jet}=4 \times 10^5$ G, $p_{\rm jet}=2.05$, $\gamma_{m, \rm jet}=3.75\times 10^2$, $\gamma_{\rm b,jet}=6.95 \times 10^2 $, $\gamma _{\rm M, jet}=2.23 \times 10^3$. As shown in Figure~\ref{fig:030329}, the $\rm SSC_{cn}$ component of the shear-accelerated electrons causes a spectral bump around 1 keV, contributing to an excess relative to the X-ray band emission from the synchrotron radiation of the shock-accelerated electron in the jet core. 

The electron distribution of the shock-accelerated electrons in the jet core derived from our analysis for the three GRBs ranges from $\sim 10^2$ to $10^3$. The acceleration timescale for shock-accelerated electrons can be estimated as \citep{1977ICRC...11..132A,1983RPPh...46..973D}
\begin{equation}
t_{\rm acc,jet}=\eta\frac{\gamma_{e,\rm jet}c}{B_{\rm jet}\Gamma_{\rm jet}\beta_{\rm sh}^{2}} ,
\end{equation}
where $\beta_{\rm sh}$ is the velocity of the internal shocks in units of $c$ and $\eta (\geq 1)$ is the gyro-factor. The maximum $\gamma_{e, \rm jet}$ is estimated through the equilibrium between acceleration and cooling timescales ($t_{\rm acc,jet} = t_{\rm cool}$, including synchrotron and SSC losses). Setting $\eta=1$, $\beta_{\rm sh}=0.5$, $\Gamma=500$, and $B_{\rm jet}=10^6$ G, the inferred $\gamma_{\rm e,M}$ is $\sim 10^6$. 
This value is three orders of magnitude larger than that derived from our analysis. The narrow $\gamma_{e, \rm jet}$ distribution and relatively low $\gamma_{\rm M, jet}$ may indicate that the residence time of electrons within the shock region is limited, preventing the full formation of an extended power-law distribution \citep{2003ApJ...589L..73L,2008ApJ...682L...5S}. Additionally, a partially ordered magnetic field configuration or underdeveloped turbulence could considerably reduce the efficiency of diffusive shock acceleration, thereby suppressing energy diffusion and resulting in a spectrally narrow electron distribution \citep{2005AIPC..801..345S,2006ApJ...645L.129L,2009ApJ...698.1523S,2013ApJ...771...54S}.

The derived $p_{\rm jet}$ is generally consistent with the prediction of the Fermi acceleration mechanism \citep{1978ApJ...221L..29B,1987ApJ...315..425K,2001MNRAS.328..393A}. 
\cite{1987ApJ...315..425K,1987ApJ...322..256K} demonstrated that the spectral index of shock-accelerated particles depends on the pitch angle distribution of the particles and the location at which the energy distribution is measured. In the high-momentum regime, the index exhibits slight variations. Their study reveals a characteristic high-energy spectrum with a slope of approximately -2.4, along with a certain degree of anisotropy in the electron distribution, particularly in relativistic shocks. These findings were corroborated by particle-in-cell (PIC) simulations, which also suggest a presence of high-energy cut-off in distribution(e.g. \citealp{2003ApJ...595..555N,2008ApJ...682L...5S,2011ApJ...739L..42H}). 
We should emphasize that our calculations in this analysis employ a simplified treatment by assuming an isotropic distribution of shock-accelerated electrons characterized as a broken power-law function as guided by the PIC simulation results 
\footnote{Most PIC simulations adopt a reduced proton-to-electron mass ratio rather than the physical value of 1836; nevertheless, the qualitative results remain largely preserved \citep{2008ApJ...673L..39S,2011ApJ...726...75S,2015ApJ...811...57A,2019MNRAS.485.5105C}. Simulations with higher mass ratios, such as those performed by \cite{2011ApJ...726...75S,2013ApJ...771...54S,2022ApJ...933...74G}, indicate that increasing the mass ratio causes a slight shift in the maximum particle energy and electron distribution peak energies. }, 
although both theoretical studies and PIC simulations suggest that these electrons can exhibit more specific spectral features and angular anisotropies, which could influence the resulting emission. 
Furthermore, the current analysis assumes a completely random magnetic field in the radiation region (e.g. \citealp{1986A&A...164L..16C}).
The presence of a partially ordered magnetic field can induce a high degree of polarization and anisotropic radiation \citep{1986A&A...164L..16C,2006ApJ...652..482P,2020MNRAS.491.3343G}. This may be potentially verified with future polarization observations \citep{2019SCPMA..6229502Z,2024JInst..19P8002K}.

The Space Variable Objects Monitor (SVOM) mission  \citep{2016arXiv161006892W,2021Galax...9..113B,2022IJMPD..3130008A} equips with a Visible Telescope (VT, R- and B- band), a Microchannel X-ray Telescope (MXT, 0.2-10 keV), ECLAIRs (ECL, 4-250 keV), and Gamma Ray Burst Monitor (GRM, 15–5000 keV). The Ground Wide Angle Camera (GWAC) covers 5400 square degrees of the sky and reaches a limit up to $V=16$ magnitude at an exposure time of 10 seconds \citep{2016arXiv161006892W}. The Einstein Probe (EP) is a space mission dedicated to discovering transients and monitoring variable objects in the X-ray band. It is equipped with the Wide-field X-ray Telescope (WXT), which provides sensitivity in the 0.5-4 keV range and features a very large instantaneous field-of-view of 3600 square degrees \citep{2015arXiv150607735Y,2022hxga.book...86Y}. Synergy observations with these telescopes in broadband spectral coverage discern emission characteristics of distinct components. This provides capacity for verifying our model. 

\begin{acknowledgements}
This work is supported by the National Key R\&D Program (2024YFA1611700) and the National Natural Science Foundation of China (grant Nos. 12203015, 12133003). This work is also supported by the Guangxi Talent Program (“Highland of Innovation Talents”).
\end{acknowledgements}

\newpage

\bibliography{sample631}{}
\bibliographystyle{aasjournal}

\end{document}